\documentclass[11pt]{amsart}

\numberwithin{equation}{section}
\begin{document}

\title[]{The total space-time of a point-mass when $\Lambda\neq0$,
and its consequences for the Lake-Roeder black hole}
\author{Leonard S. Abrams}
\address{Minimax, 24345 Crestlawn Street, Woodland Hills, CA 91367, USA}
\email{ag272@lafn.org}

\thanks{Published in Physica A {\bf 227} (1996) 131-140}
\thanks{Phone:(818)883-6298.}
\date{}

\begin{abstract}
Singularities associated with an incomplete space-time $(S)$ are
not well-defined until a boundary is attached to it. Moreover,
each boundary $(B)$ gives rise to a different singularity
structure for $S\cup B$, the resulting ``total'' space-time (TST).
Since $S$ is compatible with a variety of boundaries, it therefore
does not represent a unique universe, but instead corresponds to a
family of universes, one for each possible boundary.\par
It is shown that in the case of Weyl's space-time for a point-mass
with nonzero $\Lambda$, the boundary which he attached to it is
invalid, and when the correct one is attached, the resulting TST
is inextendible. This implies that the Lake-Roeder black hole
cannot be produced by gravitational collapse.
\end{abstract}
\maketitle

\section{Introduction}
Let $S=(M,g)$ denote an incomplete, not necessarily inextendible
space-time. It is well known \cite{ref:TCE80} that the
singularities of $S$ are not completely specified by $(M,g)$
alone, and that in order to achieve such a specification a
boundary ($B$, say) must be attached to it. [The resulting object,
$T=S\cup B$, will be termed a ``total'' space-time (TST).]\par
Moreover, thanks to the work of a dozen or so investigators in the
sixties and seventies (e.g. Refs. \cite{ref:Schmidt69},
\cite{ref:Geroch68}, \cite{ref:Dodson79}), it is now also
well-known that an incomplete space-time is compatible with a
variety of boundaries, each of which gives rise to a different
singularity structure for the resulting TST.\par Taken together,
these two facts show that an incomplete space-time by itself does
not represent a unique universe - rather, it corresponds to a
family of universes, one for each of its possible boundaries.
Changing the boundary attached to $S$ changes the universe
represented by the resulting TST \cite{ref:HE73}.\par Let us now
apply the foregoing considerations to a universe $(U_0)$
consisting of a single point-mass and a nonzero cosmological
constant $(\Lambda)$. As is well-known, the space-time $(S_0)$ for
$U_0$ was first obtained by Weyl \cite{ref:Weyl18}. Unfortunately,
Weyl's derivation (Section 2) involved a tacit assumption which is
shown here (Section 3) to be invalid. When this assumption is
eliminated (Section 4), the resulting space-time (termed
``Stavroulakis'', since its metric is a special case of one found
by him \cite{ref:Stavro81}) is isometric, and hence equivalent
\cite{ref:SW77}, to $S_0$. Thus, in this respect the invalidity of
Weyl's assumption is harmless. However, Stavroulakis' (and Weyl's)
space-time is timelike incomplete, so that it cannot represent any
universe, and thus in particular does not represent $U_0$. The
question therefore arises: What boundary must be attached to $S_0$
in order that the resulting TST represent $U_0$? Since in
Einstein's relativity point-masses are necessarily singularities
of the field, and since $U_0$ has no sources other than the single
point mass, the answer is immediate: the boundary must be a line
through the point-mass. [Equivalently, in each spatial section the
boundary must consist of a point at the location of the
point-mass.] It is here that Weyl's assumption gave rise to a
fatal flaw, since it automatically attached a boundary consisting
of a two-sphere at the location of the point-mass. (See Section
7.2.)\par It is then shown (Section 5) that with the correct
boundary, Stavroulakis' TST is inextendible and contains no black
hole. As a result (Section 6), for $\Lambda\neq 0$ no black hole
can be formed when a spherically symmetric, uncharged, nonrotating
star undergoes gravitational collapse. Lacking both a valid
derivation for a specific universe and a plausible model for its
production, it follows that the black hole found by Lake and
Roeder \cite{ref:LR77} and by Laue and Weiss \cite{ref:LW77} is
nothing more than an artifact of a historical error.\par It should
be emphasized that since a TST is a space-time-with-boundary, the
criteria for equivalence and extendibility \cite{ref:ext} are
necessarily different from those applicable to space-times.
Specifically, it follows from their definition that equivalence of
TST requires not only that their interiors be isometric, but also
that their boundaries be homeomorphic. Likewise, extendibility of
a TST requires not only that its interior be isometric to the
proper open subset of another space-time, but also that its
boundary be preserved under the mapping, i.e., that the image of
$B$ be homeomorphic to $B$ itself.

\section{Weyl's derivation}
Weyl's derivation of the metric of an uncharged point-mass when
$\Lambda\neq0$ was as follows: Starting from the most general
expression \cite{ref:Eiesland25} for the static metric which is
spherically-symmetric about the location of the point-mass (taken,
without loss of generality, to be $x=y=z=0$), viz.:
\begin{equation}\label{2.1}
g(r)=A(r)dt^2-B(r)dr^2-C(r)d\Omega^2,~~~A,B,C>0,
\end{equation}
where $r,\theta,\phi$ are quasi-spherical polar coordinates [i.e.,
$r=(x^2+y^2+z^2)^{1/2}$, etc.], he introduced a new radial
coordinate via:
\begin{equation}\label{2.2}
r^*=[C(r)]^{1/2},
\end{equation}
which transforms (\ref{2.1}) into
\begin{equation}\label{2.3}
g^*(r^*)=A^*(r^*)dt^2-B^*(r^*)dr^{*2}-r^{*2}d\Omega^2,
\end{equation}
and, of course, assigns to the location of the point-mass the
value
\begin{equation}\label{2.4}
r^*_0=[C(0+)]^{1/2}.
\end{equation}
He then solved for $A^*,B^*$ using a variational principle
equivalent to the vacuum field equations, obtaining \cite{ref:res}
\begin{equation}\label{2.5}
A^*=A^*_{\rm w}\equiv 1-\alpha/r^*-\Lambda r^{*2}/3,
\end{equation}
\begin{equation}\label{2.6}
B^*=B^*_{\rm w}\equiv 1/{A^*_{\rm w}},
\end{equation}
where $\alpha$ is a constant.\par
Unfortunately, the introduction of $r^*$ as a coordinate creates
something of a problem - since $C(0+)$ is unknown, there is no way
to determine the value of $r^*_0$, the location of the point-mass
(cf. Ref. \cite{ref:Abrams89}). This was overlooked by Weyl, who
had tacitly assumed \cite{ref:ref6} that the point-mass' location
was given by $r^*=0$. As follows from (\ref{2.4}), this can only
be true if $b^2\equiv C(0+)=0$.

\section{The invalidity of Weyl's assumption}
In order to determine whether this assumption is valid, one could
substitute (\ref{2.1}) into the vacuum field equations, solve for
$A,B,C$, and then see whether $C(0+)=0$ ia admissible (cf.
\cite{ref:Abrams89}). However, there is a simpler way - all one
need do is return to the $r$-coordinate system, in which the
location of the point-mass is known, by using (\ref{2.2}). So
doing changes (\ref{2.3}) into (\ref{2.1}), and (\ref{2.5}) and
(\ref{2.6}) into:
\begin{equation}\label{3.1}
A=1-\alpha/{\sqrt C}-\Lambda C/3,
\end{equation}
\begin{equation}\label{3.2}
B={C'}^2/{(4AC)},
\end{equation}
respectively, where at this stage $C$ is any positive analytic
function of $r$ for $r>0$.\par
It follows by inspection of (\ref{3.2}) that in order to insure
the positivity of $B$, it is necessary and sufficient that
${C'}^2$ be nonvanishing for $r>0$. This in turn requires that
either $C'>0$ for $r>0$ or $C'<0$ for $r>0$. Since the metric must
tend to that of Schwarzschild as $\Lambda\rightarrow 0$, the only
possible choice is:
\begin{equation}\label{3.3}
C'>0~~{\rm for}~~r>0.
\end{equation}
The constraints on $C$ required to assure the positivity of $A$
cannot be determined by mere inspection of (\ref{3.1}), since the
behavior of a cubic is involved. The analysis is relegated to
Appendix A, and only the results are given here:
\begin{eqnarray}
&{\rm For}~~ \Lambda<0 &C(0+)\equiv b^2\geq C_0>0,\label{3.4}\\
&{\rm For}~~ 0<\Lambda<\Lambda_0 &C_3\geq b^2\geq C_2>0
~~[\Lambda_0\equiv 4/{(9\alpha^2)}],\label{3.5}\\
&{\rm For}~~ \Lambda_0\leq\Lambda &{\rm No~metrics~with}~A~
{\rm satisfying~(3.1)~exist.}\nonumber
\end{eqnarray}
(The values of $C_0,C_2$ and $C_3$ are given in Appendix A.) In
view of the foregoing, it follows that the necessity that $A$ be
positive for $r>0$ requires that $b^2>0$, which renders Weyl's
assumption invalid for all admissible values of $\Lambda$.

\section{Additional restrictions on $C$}
As shown by Doughty \cite{ref:Doughty81}, the locally measured
acceleration of an uncharged test particle in a gravitational
field with metric given by (\ref{2.1}) is
\begin{equation}\label{4.1}
a=|A'|/{(2A\sqrt B)}.
\end{equation}
Substituting from (\ref{3.1}) and (\ref{3.2}), this becomes
\begin{equation}\label{4.2}
a=|\alpha/{(2C^{3/2})}-\Lambda/3|(C/A)^{1/2},
\end{equation}
which in turn tends to
\begin{eqnarray}
a_0=|\alpha/{(2b^3)}-\Lambda/3|b/{\sqrt{A(0+)}}
& {\rm as} &r\rightarrow 0.\label{4.3}
\end{eqnarray}
As pointed out in Ref. \cite{ref:Abrams89}, the value of $a_0$ is
a scalar differential invariant of the space-time. Thus, different
values of $a_0$ give rise to inequivalent space-times. Here, as in
Ref. \cite{ref:Abrams89}, we shall take $a_0$ to have its
Newtonian value, infinity. In view of the nonzero values of $b$
found to be required in Section 3, it follows from (\ref{4.3})
that the only way to make $a_0$ infinite is to choose $A(0+)=0$.
As indicated in Appendix A, this in turn requires that:
\begin{eqnarray}
b=&\sqrt{C_0} &{\rm for}~~\Lambda<0, \label{4.4}\\
=&\sqrt{C_2} &{\rm for}~~0<\Lambda<\Lambda_0.\label{4.5}
\end{eqnarray}
The final restriction on $C$ is obtained by considering the
limiting behavior of the point-mass space-time as
$r\rightarrow\infty$. In the $\Lambda=0$ case, this behavior was
determined by requiring that as the proper distance from the
point-mass became unboundedly large, the metric must approach that
in which no point-mass is present - i.e., Minkowski's. Here,
however, in each spatial section ($t=$ constant), the integral
\begin{equation}\label{4.6}
R_{\rm max}\equiv{\int_0^{\alpha}}{{\sqrt B}dr}
={1/2}{\int_{b^2}^{k^2}(1/{\sqrt{AC})dC}}
\end{equation}
may be either finite or infinite, depending on whether $k$
[$\equiv\sqrt{C(\infty)}$] is finite or infinite. In the former
case, which is always true when $0<\Lambda<\Lambda_0$ (see
Appendix A), there are no events which are at infinite proper
distances from the point-mass, so that no asymptotic condition
corresponding to that for $\Lambda=0$ exists. However, it is easy
to show (Appendix B) that if $k<k_{\rm max}\equiv\sqrt{C_3}$, then
the associated TST is extendible to one for which $k=k_{\rm max}$,
so that for $0<\Lambda<\Lambda_0$ the only viable choice of $k$ is
\begin{equation}\label{4.7}
\sqrt{C(\infty)}=k_{\rm max}.
\end{equation}
Similarly, if for $\Lambda<0$ the limiting value of $C$ as
$r\rightarrow\infty$ were taken to be finite, then the associated
TST would be extendible to one for which $C(\infty)$ is infinite
(see Appendix B). Thus, for $\Lambda<0$ the only viable choice of
$k$ is given by
\begin{equation}\label{4.8}
\sqrt{C(\infty)}=\infty.
\end{equation}
In this case, since $R_{\rm max}$ is infinite, then for large $r$
the influence of the point-pass on the space-time geometry must
necessarily vanish, just as in the $\Lambda=0$ case. Thus, the
metric must approach that for which no point-mass is present.
Synge has shown \cite{ref:Synge60} that this limiting metric is
one of constant curvature, namely the anti-de Sitter metric
\begin{equation}\label{4.9}
g_{\rm ads}=(1-\Lambda r^2/3)dt^2-(1-\Lambda r^2/3)^{-1}dr^2
-r^2d\Omega^2.
\end{equation}
Comparison of this with (\ref{3.1}) and (\ref{3.2}) shows that for
this case ($\Lambda<0$) it is necessary that
\begin{eqnarray}
C/{r^2}\rightarrow 1 &{\rm as} & r\rightarrow\infty.\label{4.10}
\end{eqnarray}
A suitable $C$ for this case is thus
\begin{equation}\label{4.11}
C_{\rm inf}\equiv (r+b)^2,
\end{equation}
with $b$ given by (\ref{4.4}). However, even knowing that
$g\rightarrow g_{\rm ads}$ does not suffice to determine $\alpha$,
because the presence of the $\Lambda(r+b)^2$ term in (\ref{3.1})
makes the space-time more-and-more non-Newtonian as
$r\rightarrow\infty$, and thus the Kepler-orbit requirement for
distant test particles cannot be invoked to identify $\alpha$ with
$2m$.\par
Incidentally, note that the strict monotonicity required of $C$ by
virtue of (\ref{3.3}), together with the limiting values of $C$,
shows that (\ref{2.2}) is a diffeomorphism, so that the
space-times whose $A,B$ are given by (\ref{3.1}) and (\ref{3.2})
are isometric to Weyl's. Thus the only difference between Weyl's
TST and those found here lies in the difference between their
boundaries (see Section 7.2, below).\par

\section{Singularities of the total space-times}

In Ref. \cite{ref:LR77} the Kretschmann invariant $f\equiv
R_{ijkm}R^{ijkm}$ is calculated for Weyl's metric. Transforming
this via (\ref{2.2}) and letting $r\rightarrow 0$ shows that $f$
approaches
\begin{equation}
f_0\equiv 12\alpha^2/{b^2}+24\Lambda^2/9,
\end{equation}
so that for the values of $b$ found to be required in the previous
Section, $f$ is bounded as $r\rightarrow 0$. Since all other
scalar differential invariants are functions of $f$, it follows
that there are no curvature singularities at the location of the
point-mass.\par However, (\ref{2.1}) shows that the proper
circumference of the circle $r=\varepsilon$ tends to $2\pi b>0$ as
$\varepsilon\downarrow 0$, while the proper radius of that circle
is easily seen [from (\ref{2.1}) and (\ref{3.2})] to tend to zero
in that limit. These results are coordinate-independent: once a
metric has been brought into the form of (\ref{2.1}), the only
transformations which preserve the form of (\ref{2.1}) are
$t=K\bar{t}+q$ ($K\neq 0$), $r=h(\bar{r})$ ($h, h^{-1}\in
C^{\omega}$), neither of which alters the proper radius or proper
circumference of $r=\varepsilon$.\par Since the boundary at $r=0$
is necessarily a point in each spatial section (see Section 1),
these properties of the radius and circumference of
$r=\varepsilon$ constitute a violation of elementary flatness at
$r=0$, and a fortiori a quasiregular singularity \cite{ref:ES77}
which renders these TSTs inextendible.

\section{Gravitational collapse}

As pointed out in Ref. \cite{ref:Abrams89}, the metric
representing the exterior of a spherically symmetric star
undergoing gravitational collapse to a point is subject to the
same requirements as that of a point-mass except for those
relating to its behavior at $r=0$. It thus has the same form as
that of a point-mass [i.e., (\ref{2.1}) cum (\ref{3.1}) and
(\ref{3.2})], but different parameter values since $C$ need only
make $A$ positive for all $r>r_b$, where $r_b$ denotes the radial
coordinate of the star's boundary. As $r_b\rightarrow 0$, this
assures the positivity of $A$ for all $r>0$, so that no horizon,
and a fortiori no black hole, can be formed at any stage of the
collapse. Thus, just as in the case where $\Lambda=0$, elimination
of the invalid assumption regarding the location of the point-mass
also deprives the Lake-Roeder black hole of the only mechanism
suggested for its production.

\section{Summary and discussion}

\subsection{}
    The only physically sensible metrics for the point-mass when
$\Lambda\neq0$ are given by
\begin{eqnarray}\nonumber
g=Adt^2-Bdr^2-Cd\Omega^2,\\\nonumber
A=1-\alpha/{\sqrt{C}}-\Lambda C/3, &B={C'}^2/{(4AC)}.\\\nonumber
\end{eqnarray}
The form of $C$ depends on $\Lambda$:\bigskip
\subsubsection{$\Lambda<0$}
~\par
$C=(r+b)^2$ with $b$ given by (\ref{4.4}). Any positive, analytic,
strictly monotonic increasing function of $r$ having the same
value of $b$ and tending to infinity like $r^2$ can be used in
place of this $C$ and, with the space-time's boundary taken to be
a line through the source, will give rise to an equivalent TST. The
value of $\alpha$ must be determined from other
considerations.\bigskip
\subsubsection{$0<\Lambda<\Lambda_0$}
~\par
$C$ is any analytic, strictly monotonic increasing function of $r$
satisfying $C(0+)=b^2$ and $C(\infty)=k^2_{\rm max}$, where $b$ is
given by (\ref{4.5}) and $k_{\rm max}=\sqrt{C_3}$. With the
space-time's boundary taken to be a line through the source, all
such $C$ give rise to equivalent TSTs. The value of $\alpha$ must
be determined from other considerations.

\subsection{}
As noted in Section 1, the $r=0$ boundary of the point-mass TSTs
associated with the above metrics must necessarily be a point in
each spatial section. Under the transformation $r^*=[C(r)]^{1/2}$
this boundary becomes $r^*=[C(0+)]^{1/2}=b$. That is to say, the
locus $r^*=b$ is a point in each spatial section. But Weyl's
assumption that the origin of the $x,y,z$ coordinates is located
at $r^*=0$ automatically makes $r^*=b$ a two-sphere in each
spatial section of his space-time. Thus, Weyl's assumption led to
the attachment of a two-sphere, rather than a point, as the
boundary of each spatial section of his space-time at $r^*=b$ -
i.e., at the actual location of the point-mass. Consequently, his
TST has a different singularity structure than that associated
with the space-times having the above metrics. Since the metrics'
derivation shows that the TSTs obtained here are the only possible
ones for the point-mass when $\Lambda\neq 0$, it follows that his
TST does not represent such an object. A fortiori, the same is
true of the black hole-containing analytic extension of Weyl's TST
found in 1977 by Lake and Roeder \cite{ref:LR77} and,
independently, by Laue and Weiss \cite{ref:LW77} (to say nothing
of the fact that this extension alters the topology of the
boundary of Weyl's TST, and thus represents a different universe
than the latter). This in turn means that this extension cannot
serve as the limiting space-time of a spherically symmetric star
undergoing collapse to a point. Thus, it is impossible to produce
this extension by gravitational collapse. Lacking both a
theoretical basis (i.e. a valid derivation from a set of
postulates characterizing a specific universe) and a mechanism for
its production, it follows that this extension, and with it its
black hole, are merely artifacts of Weyl's error.

\subsection{}
The relation between $\alpha$ and $m$ is still open.
\section{Acknowledgments}
It is a pleasure to acknowledge helpful correspondence with G.F.R.
Ellis and R. Geroch, as well as numerous conversations with B.
O'Neill and R. Greene. Such errors as remain are mine alone.

\appendix
\section{Constraints on $C$ required to make $A$ positive}
It is expedient to consider the following three intervals of
$\Lambda$: $\Lambda<0$; $0<\Lambda<\Lambda_0$; $\Lambda_0\leq\Lambda$,
where $\Lambda_0=4/9\alpha^2$.
\subsection{$\Lambda<0$}
It is shown in Ref. \cite{ref:LR77} (with $r$, $2m$ there replaced
by $\sqrt{C}$, $\alpha$ respectively) that $A$ is positive iff
$C>C_0$, and vanishes only at $C=C_0$, where:
\begin{equation}\label{A.1}
C_0\equiv (\mu+\sqrt{\mu^2-\Lambda^{-3}} )^{1/3}
+(\mu-\sqrt{\mu^2-\Lambda^{-3}} )^{1/3}>0,
\end{equation}
in which
\begin{equation}\label{A.2}
\mu\equiv 3\alpha/{(-2\Lambda)}>0.
\end{equation}
Together with (\ref{3.3}), the foregoing shows that in order to
make $A>0$, it is necessary and sufficient that
\begin{equation}\label{A.3}
\sqrt{C(0+)}\equiv b\geq \sqrt{C_0}>0.
\end{equation}
\subsection{$0<\Lambda<\Lambda_0$}
In this case, it is shown in Ref. \cite{ref:LR77} that $A$ is
positive iff:
\begin{equation}\label{A.4}
C_2<C<C_3,
\end{equation}
and vanishes only when $C=C_2$ or when $C=C_3$, where:
\begin{equation}\label{A.5}
\sqrt{C_2}\equiv(2/{\sqrt{\Lambda}})\cos{(\xi/3)}>0,
\end{equation}
\begin{equation}\label{A.6}
\sqrt{C_3}\equiv(2/{\sqrt{\Lambda}})\cos{(\xi/3+4\pi/3)},
\end{equation}
\begin{equation}\label{A.7}
\cos{(\xi)}\equiv-3\alpha\sqrt{\Lambda}/2,~~\pi<\xi<3\pi/2,
\end{equation}
In combination with (\ref{3.3}), the left-hand-most inequality in
(\ref{A.4}) requires that
\begin{equation}\label{A.8}
b\geq\sqrt{C_2}>0.
\end{equation}
\section{Extendibility of total space-times}
Consider first the case $\Lambda<0$ with $k$ finite, so that the
metric ($g_{\rm fin}$) is given by (\ref{2.1}), (\ref{3.1}),
(\ref{3.2}) with $C(r)\equiv C_{\rm fin}(r)$, where
$C_{\rm fin}(\infty)=k^2$. Consider also the same metric, but with
$C(r)\equiv C_{\rm inf}(r)$, where $C_{\rm inf}(\infty)=\infty$,
and let this second metric be designated by $g_{\rm inf}$.\par
Thanks to the strict monotonicity of both $C'$s, the mapping
$\zeta:M_0\rightarrow \zeta(M_0)$ by
$C_{\rm fin}(r)=C_{\rm inf}(\bar{r})$ is an analytic
diffeomorphism. It is also an isometry to the open proper
submanifold $\bar{r}<C_{\rm fin}^{-1}(k^2)$ of ($M_0,g_{\rm
inf}$). Finally, the boundary that is attached to
($M_0,g_{\rm fin}$) at $r=\infty$ must be a two-sphere in each
spatial section, rather than a point, since the latter would give
rise to a second point-like singularity, which is inconsistent
with the single point-mass hypothesis. Thus the space-time
($M_0,g_{\rm fin}$) is isometric to a proper open submanifold of
($M_0,g_{\rm inf}$), and its boundary is homeomorphic to its
image. This means that the TST of ($M_0,g_{\rm fin}$) is
extendible to that of ($M_0,g_{\rm inf}$).\par
Consider next the case $0<\Lambda<\Lambda_0$. In this case the
metric ($g_k$) is given by (\ref{2.1}), (\ref{3.1}),
(\ref{3.2}) with $C(r)\equiv C_k(r)$, where $C_k(0+)=C_2$ and
$C_k(\infty)=k^2<k^2_{\rm max}\equiv C_3$. Consider also the same
metric, but with $C\equiv C_{\rm max}(r)$, where
$C_{\rm max}(\infty)=k^2_{\rm max}$, and let this second metric be
denoted by $g_{\rm max}$. Replacing fin by $k$ and inf by max in
the above proof of the extendibility of the TST of
($M_0,g_{\rm fin}$), the same proof is seen to hold for this case
as well, so that the TST of ($M_0,g_k$) is extendible to that of
($M_0,g_{\rm max}$).

\bibliographystyle{amsplain}

\begin{thebibliography}{}
\bibitem{ref:ext} By ``extendibility'' here will always be meant
analytic extendibility.
\bibitem{ref:ref6} Ref. [6], p. 226: ``Sie entspricht dem Fall,
dass um das Zentrum ein Massenkugel liegt.'' (``This corresponds
to the case, in which a spherical mass lies at the center.'').
\end{thebibliography}

\end{document}